\documentclass[letterpaper, 10 pt, conference]{ieeeconf}  %

\IEEEoverridecommandlockouts                              %

\usepackage{graphicx}
\usepackage{ascmac}

\usepackage{bm}
\usepackage{amsmath}
\usepackage{amssymb}
\usepackage{amsfonts}
\usepackage{color}
\usepackage{wrapfig}
\usepackage{url}
\usepackage{cite}
\usepackage{siunitx}
\usepackage{color}
\usepackage{comment}

\usepackage{fancyhdr}

\title{\LARGE \bf
Frequency response of diffusion-based molecular communication channels in bounded environment*
}

\author{Taishi Kotsuka$^{1}$ and Yutaka Hori$^{1}$%
\thanks{*This work was supported in part by JSPS KAKENHI Grant Number 21H05889, JST SPRING Grant Number JPMJSP2123, and the Keio University Doctorate Student Grant-in-Aid Program from Ushioda Memorial Fund.}%
\thanks{$^{1}$T. Kotsuka and Y. Hori are with the Department
of Applied Physics and Physico-Informatics, Keio University, Kanagawa 223-8522 Japan. Correspondence should be addressed to Y. Hori. {\tt yhori@appi.keio.ac.jp}}%
}

\fancypagestyle{firstpage}{\lfoot{\scriptsize \copyright 2022 IEEE. Personal use of this material is permitted. Permission from IEEE must be obtained for all other uses, in any current or future media, including reprinting/republishing this material for advertising or promotional purposes, creating new collective works, for resale or redistribution to servers or lists, or reuse of any copyrighted component of this work in other works.} \cfoot{}}

\begin{document}

\maketitle
\thispagestyle{firstpage}

\begin{abstract}

Recently, molecular communication (MC) has been studied as a micro-scale communication between cells or molecular robots. In previous works, the MC channels in unbounded environment was analyzed. However, many of the experimentally implemented MC channels are surrounded by walls, thus the boundary condition should be explicitly considered to analyze the dynamics of MC channels. In this paper, we propose a framework to analyze the frequency response of one-dimensional MC channels based on a diffusion equation with a boundary. In particular, we decompose the MC channel into the diffusion system and the boundary system, and show the relation between the cut-off frequency of the MC channel and the communication distance based on the transfer function. We then analyze the frequency response of a specific MC channel and reveal that the boundary can restrict the communication bandwidth of the MC channel.

\end{abstract}

\bibliographystyle{IEEEtran}

\section{Introduction}
In nature, bacteria are known to have a molecular communication (MC) mechanism that controls the behavior of cell populations through signal molecules that can diffuse between cells \cite{Teresa2000}. In recent years, the MC has been studied as a means of micro-scale communication between cells or molecular robots \cite{tatsuya2018molecular,Farsad2016}. Especially, by forming feedback between cells or molecular robots via MC, which is called as MC systems, it would be possible to stabilize the work of individual cells and suppress the effects of noise, thereby achieving cooperative work that cannot be achieved by a single cell. The MC systems are expected to have practical applications such as target tracking, drug delivery and real-time health monitoring \cite{Femminella2015,Gao2014,sldner2019survey,Din2016,Chude-Okonkwo2017}. In \cite{Din2016}, the potential of bacteria with the MC mechanism as a drug delivery platform was demonstrated. The engineered bacteria share information using diffusion of the signalling molecules to release drug efficiently in synchronized cycles. 

\smallskip
\par
For the practical application of MC systems, it is important to systematically develop analysis and design methods for MC systems. In control engineering, there are mainly two types of models for MC systems. One is the MC systems based on a reaction-diffusion model that assumes that cells are in close contact with each other. To date, several control analysis of the reaction-diffusion based MC systems were conducted \cite{Kotsuka2019a,yutaka2015coordinated}. The other model for MC systems is the decomposed model into intracellular systems and MC channels. Since the decomposed models can take the distance between cells into account, they are versatile in terms of applications. Control theory for intracellular biomolecular circuits was extensively studied recently \cite{DelVecchio2016,Prescott2014,Reeves2019}. However, there are not many studies on the systematical approach to design MC channels, thus it is necessary to develop control-theoretic studies of communication channels toward versatile practical application. 

\smallskip
\par
To construct control theory to design MC channels, the basic characteristics of MC channels such as frequency response needs to be analyzed. To date, there are only a few works that proposed system theoretic methods for MC channels \cite{Pierobon2010,Pierobon2014,Chude-Okonkwo2015}. In \cite{Pierobon2010}, the frequency response of the MC channel was analyzed based on a diffusion model of signal molecules in unbounded 3D environment. This model is suitable for MC in a large space, but not for MC channels surrounded by membranes or walls, such as the channel developed in \cite{Dupin2019}. Since bounded diffusion models are more complicated than unbounded models, one of the challenges is the system-theoretical analysis of MC channels in bounded environment. 

\smallskip
\par
In this paper, we propose a framework to analyze the frequency response of one-dimensional MC channels based on a diffusion equation with a boundary condition determined by the dynamics of membrane transport and its flux. The proposed method enables comparing the cut-off frequencies of the decomposed subsystems that have physical roles to identify which subsystem determines the communication bandwidth of the entire MC channel. Using the proposed approach, we show the relation between the communication distance and the cut-off frequency determined by the dynamics at the boundary. We further analyze the frequency response of the MC channel with the passive transport of specific molecules in the membrane to reveal that the dynamics at the boundary can restrict the bandwidth of the MC channel depending on the communication distance. This numerical example illustrates the importance to consider the dynamics at the boundary to analyze and design MC systems.

\smallskip
\par
This paper is organized as follows. In the next section, we introduce and model a one-dimensional MC system consisting of the diffusion layer and the boundary layer. In Section \ref{sec:frequency}, we decompose the MC channel into two subsystems based on the transfer functions. We then show the limitation of the communication distance by the cut-off frequency determined by the dynamics at the boundary. The frequency response of a specific MC system with the passive membrane transport in Section \ref{sec:numerical} based on the decomposed subsystems. Finally, the paper is concluded in Section \ref{conc-sec}.

\section{Modeling of a MC channel}
\label{sec:modeling}

\begin{figure}
    \centering
    \includegraphics[width=0.9\linewidth]{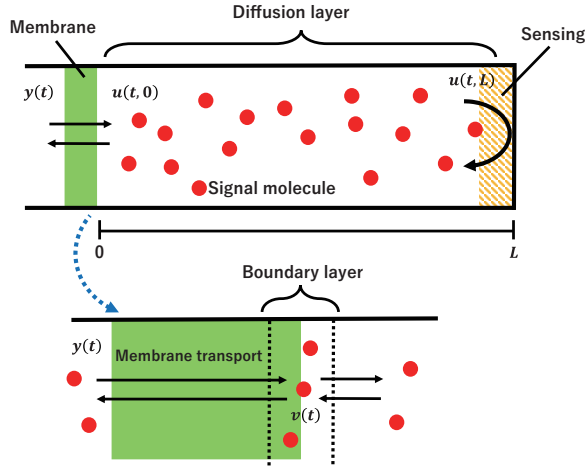}
    \caption{The molecular communication channel with the boundary layer.}
    \label{fig:molecularcommun}
\end{figure}

We consider a one-dimensional MC channel, where signal molecules transmitted from the left end of the channel diffuse in a fluidic medium [$0,L$] with $L$ being the communication distance. At the right end of the channel, the intensity of the signal is detected as the concentration of the molecule as shown in Fig. \ref{fig:molecularcommun}. The MC channel consists of two parts: the diffusion layer and the boundary layer. 

\smallskip
\par
In the diffusion layer, the signal molecules move in a fluidic medium according to the Fick’s first law, the flux $J(t,r)$ at time $t$ and location $r$ can be calculated by the spatial gradient of the concentration of the signal molecule $u(t,r)$ as
\begin{eqnarray}
J(t,r) = -\mu \nabla u(t,r),
\label{eq:flux}
\end{eqnarray}
where $\nabla$ is the Laplace operator, and $\mu$ is the diffusion coefficient, which is a constant that does not depend on the molecular concentration. Substituting the Fick's first law (\ref{eq:flux}) into the continuity equation
\begin{eqnarray}
\frac{\partial u(t,r)}{\partial t} = -\nabla J(t,r),
\label{eq:continuity}
\end{eqnarray}
we obtain the diffusion equation as
\begin{eqnarray}
\frac{\partial u(t,r)}{\partial t} &=& \mu\frac{\partial^2 u(t,r)}{\partial r^2}.\label{eq:dif}
\end{eqnarray}
Moreover, we set the initial condition as
\begin{eqnarray}
u(0,r) &=& 0.\label{eq:initial}
\end{eqnarray}

\par
At the boundary of the receiver side, the molecular concentration $u(t,L)$ reached at the boundary is sensed as the signal intensity, and then the information transmits to the downstream. Since the sensed molecules at the boundary bounce off the wall and diffuse in the medium again, the Neumann boundary condition is applied at the boundary of the receiver side as
\begin{eqnarray}
\frac{\partial u(t,L)}{\partial r} = 0.
\label{eq:neumann}
\end{eqnarray}

\smallskip
\par
In the boundary layer, there are two input sources: one comes from the upstream via a membrane or a filter, and the other flows from the diffusion layer. There are several ways for molecules to cross the membrane, and in some cases they are secreted by interacting with various other molecules in the membrane. Therefore, the dynamics of the molecular concentration $v(t)$ in the boundary layer can be modeled by
\begin{eqnarray}
\frac{d\bm{x}(t)}{dt} &=& \bm{f}\left(\bm{x}(t),\bm{z}(t)\right), \label{eq:blayer}\\
v(t) &=& h(\bm{x}(t),\bm{z}(t)),\nonumber
\end{eqnarray}
where $\bm{x}(t)$ represents the concentration of molecules associated with the membrane transport. The vector $\bm{z}(t) = [y(t), \partial u(t,0)/\partial r]^T$ is an input, where $y(t)$ is the molecular concentration in the upstream and $\partial u(t,0)/\partial r$ is the flux from the diffusion layer. The vector $\bm{f}(\cdot)$ is a function representing the dynamics of the membrane transport system, and the function $h(\cdot)$ expresses the output and the feedthrough term. Since the boundary layer corresponds to the left boundary of the diffusion layer, we apply the Dirichlet boundary condition as $u(t,0)=v(t)$ for the diffusion equation (\ref{eq:dif}). 

\par
\smallskip
\noindent
{\bf Example.} We consider the case, in which the membrane transport system is a passive transport of specific molecules depending on the difference of the concentration between inside and outside of the transmitter cell, and the molecules are sensed by receptors of the receiver cell at the right end. The dynamics (\ref{eq:blayer}) can be rewritten as
\begin{eqnarray}
\frac{d v(t)}{d t} = k\left(y(t)-v(t)\right) + \hat{\mu}\frac{\partial u(t,0)}{\partial r},
\label{eq:emission}
\end{eqnarray}
where $k$ is the rate constant of molecular transport through the membrane, and $\hat{\mu}$ is the diffusion velocity. Here, $x(t)=v(t)$, $f(\cdot)$ equals to the right hand side of Eq. (\ref{eq:emission}), and $h(\cdot)=x(t)$. 
At the right end, the information transmits to the receiver cell, by which the molecules bind to the receptors. Since the molecules can be released from the receiver, we can apply the Neumann boundary condition in Eq. (\ref{eq:neumann}). 

\smallskip
\par
In the diffusion channel without a membrane, such as conventional heat diffusion channels, the communication bandwidth is limited only by the diffusion system (\ref{eq:dif}). However, in the MC channel, the dynamics in the boundary layer may affect the bandwidth and the communication distance. In what follow, we analyze the frequency response of the MC channel with the boundary layer to fully characterize the input-output relation of the MC channel. To this end, we derive the transfer function of the MC channel. We then reveal the relationship between the dynamics in the boundary layer and the bandwidth in which signals can be transmitted in the MC channel.
 
\section{Frequency response analysis of the MC channel}
\label{sec:frequency}
 
In this section, we analyze the frequency response of the MC channel based on the transfer functions. We then show the limitation of the communication distance by the cut-off frequency determined by the dynamics in the boundary layer.
 
\subsection{Transfer characteristics of the MC channel}
\label{sec:transfer}

\begin{figure}
    \centering
    \includegraphics[width=0.85\linewidth]{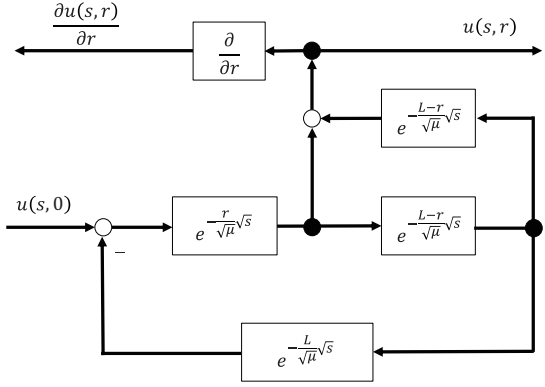}
    \caption{The block diagram of the diffusion process.}
    \label{fig:dblock}
\end{figure}

We first derive the transfer function of the diffusion system (\ref{eq:dif}). We take Laplace transform for the time variable $t$ and the spatial variable $r$ to obtain the transfer function (see Appendix \ref{sec:dtf}). Specifically, we obtain the molecular concentration and the flux at the location $r$ in $s$ domain as
\begin{eqnarray}
u(s,r) &=& G_D(s,r)u(s,0),\label{eq:rtrf}\\
\frac{\partial u(s,r)}{\partial r} &=& G_F(s,r)u(s,0),
\label{eq:rtrfder}
\end{eqnarray}
where
\begin{eqnarray}
G_D(s,r) &=& \frac{e^{-\frac{r}{\sqrt{\mu}}\sqrt{s}}+e^{\frac{r-2L}{\sqrt{\mu}}\sqrt{s}}}{1+e^{-2\frac{L}{\sqrt{\mu}}\sqrt{s}}},\label{eq:gdr}\\
G_F(s,r) &=& \sqrt{\frac{s}{\mu}}\frac{-e^{-\frac{r}{\sqrt{\mu}}\sqrt{s}}+e^{\frac{r-2L}{\sqrt{\mu}}\sqrt{s}}}{1+e^{-2\frac{L}{\sqrt{\mu}}\sqrt{s}}}.\label{eq:gfr}
\end{eqnarray}
Note that $G_D(s,r)$ and $G_F(s,r)$ can be decomposed into the feedback connection of exponential transfer functions as shown in Fig. \ref{fig:dblock}, where $e^{-R\sqrt{s}/\sqrt{\mu}}$ represents a transfer function of diffusion for distance $R$ when the right boundary exists at infinity with the boundary condition $u(t,\infty)=0$. The structure of Fig. \ref{fig:dblock} indicates the physical phenomenon of diffusion, such as the feedback loop by the exponential transfer function represents that the signal reaching location L bounces off the wall. 

\smallskip
\par
Since the molecular concentration at $r=L$ is sensed as signal intensity and the flux at $r=0$ inputs into the boundary layer in the MC channel shown in Fig. \ref{fig:molecularcommun}, we need the diffusion transfer function to $r=L$ and the flux transfer function to $r=0$, that is,
\begin{eqnarray}
G_D(s,L) &=& \frac{2e^{-\frac{L}{\sqrt{\mu}}\sqrt{s}}}{1+e^{-2\frac{L}{\sqrt{\mu}}\sqrt{s}}},\label{eq:g1}\\
G_F(s,0) &=& \sqrt{\frac{s}{\mu}}\frac{-1+e^{-2\frac{L}{\sqrt{\mu}}\sqrt{s}}}{1+e^{-2\frac{L}{\sqrt{\mu}}\sqrt{s}}}\label{eq:g2}.
\end{eqnarray}

\smallskip
\par
Next, we consider the transfer functions of the dynamics in the boundary layer. We linearize the system (\ref{eq:blayer}) as
\begin{eqnarray}
v(s) = G_1(s)y(s) + G_2(s)\frac{\partial u(s,0)}{\partial r},\label{eq:blsys}
\end{eqnarray}
where $G_1(s)$ is the transmission system that outputs the molecular concentration in the boundary layer for the input from the molecular concentration in the upstream system, and $G_2(s)$ is the system that outputs the molecular concentration in the boundary layer for the input from the concentration gradient at the left boundary of the diffusion layer. 

\smallskip
\par
\begin{figure}
     \centering
     \includegraphics[width=0.99\linewidth]{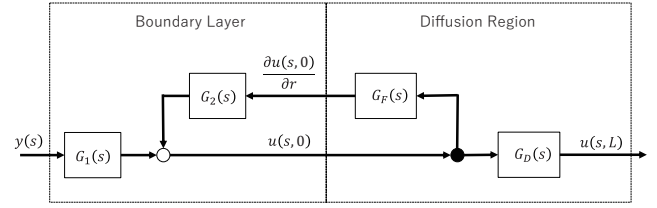}
     \caption{The block diagram of the molecular communication channel.}
     \label{fig:wblock}
 \end{figure}

Putting the diffusion layer and the boundary layer together, the block diagram of the whole MC channel can be expressed as shown in Fig. \ref{fig:wblock}. Note that the system without a membrane assumes $G_1(s)=1$ and $G_2(s)=0$ since the input $y(s)$ directly determines the value $u(s,0)$ at $r=0$. 

\subsection{Limitation of communication distance by bandwidth}
\label{sec:limit}

We analyze the frequency response of the MC channel based on the transfer functions. First, we decompose the MC channel into subsystems and clarify the role of each subsystem in the channel. We then show the limitation of the communication distance by the cut-off frequency of the decomposed subsystems.

\smallskip
\par
We decompose the MC channel shown in Fig. \ref{fig:wblock} into two subsystems: the diffusion system $G_D(s,L)$ and the boundary system $G_B(s)$ as
\begin{eqnarray}
u(s,L) = G_D(s,L)G_{B}(s) y(s),\label{eq:3div}
\end{eqnarray}
where
\begin{eqnarray}
G_{B}(s) &=& \frac{G_1(s)}{1-G_2(s)G_F(s,0)}\label{eq:GDB}
\end{eqnarray}
is the transfer function representing the dynamics of the membrane transport of signal molecules. It should be noted that the existence of the boundary layer, $G_B(s)$, is the unique feature of the MC channel, since $G_B(s) = 1$ when the signal molecules from the upstream system are directly input to the diffusion layer, \textit{i.e.} $u(t,0) = y(t)$. Therefore, we here aim to analyze the impact of the boundary system $G_B(s)$ on the communication bandwidth of the MC channel. 

\smallskip
\par
We consider the cut-off frequency of the MC channel to analyze the communication bandwidth. The cut-off frequency of the MC channel is dominated by the smallest cut-off frequency in the subsystems $G_B(s)$ and $G_D(s,L)$. Thus, comparing the cut-off frequencies of the decomposed subsystems, we can find which subsystem restricts the bandwidth of the MC channel. We define the cut-off frequency as the smallest frequency, where the gain reaches -6 dB. In what follows, we find the cut-off frequency of the diffusion system $\omega_D$. We first show that the gain of the diffusion transfer function $G_D(s,L)$ is a monotonically decreasing function. We then find the relation between the cut-off frequency of $G_D(s,L)$ and the communication distance $L$ using binary search.

\par
\smallskip
\noindent
{\bf Proposition 1.} Consider the diffusion transfer function $G_D(s,L)$ in Eq. (\ref{eq:g1}). The gain of the diffusion transfer function 
\begin{eqnarray}
|G_D(j\omega,L)| = \frac{2e^{-\hat{\omega}}}{(1+2e^{-2\hat{\omega}}\cos{(2\hat{\omega})}+e^{-4\hat{\omega}})^{\frac{1}{2}}}\label{eq:prop}
\end{eqnarray}
is a monotonically decreasing function for $\hat{\omega}>0$, where $\hat{\omega} = L\sqrt{\omega}/\sqrt{2\mu}$. 

\smallskip
\par
\noindent
{\bf Proof.} We consider the derivative of the gain of the diffusion transfer function as
\begin{eqnarray}
\frac{d|G_D(j\omega,L)|}{d\hat{\omega}} = \frac{2e^{-\hat{\omega}}(-1+2e^{-2\hat{\omega}}\sin{2\hat{\omega}}+e^{-4\hat{\omega}})}{H(\hat{\omega})},\label{eq:gaind}
\end{eqnarray}
where
\begin{eqnarray}
H(\hat{\omega}) = ((1+e^{-2\hat{\omega}}\cos{(2\hat{\omega})})^2 + e^{-4\hat{\omega}}\sin^2{2\hat{\omega}})^{\frac{3}{2}}.\label{eq:denom}
\end{eqnarray}
Since $H(\hat{\omega})\geq0$, we show 
\begin{eqnarray}
-1+2e^{-2\hat{\omega}}\sin{2\hat{\omega}}+e^{-4\hat{\omega}}<0\label{eq:state2}
\end{eqnarray}
for all $\hat{\omega}\geq0$. Eq. (\ref{eq:state2}) is equivalent to
\begin{eqnarray}
F(\hat{\omega}) := e^{-4\hat{\omega}} + 2e^{-2\hat{\omega}}\sin{2\hat{\omega}}<1 \quad\text{for $\forall \hat{\omega}>0$.}\label{eq:state3}
\end{eqnarray}
$F(\hat{\omega})$ can be bounded from above as
\begin{eqnarray}
F(\hat{\omega}) \leq e^{-\pi} + 2e^{-\frac{\pi}{2}} < 1
\end{eqnarray}
for $\hat{\omega}>\pi/4$. Thus, the statement (\ref{eq:state3}) is not violated if it is not for $0<\hat{\omega}\leq\pi/4$. In what follows, we consider the range $0<\hat{\omega}\leq\pi/4$. We take the derivative of the function $F(\hat{\omega})$ as
\begin{eqnarray}
\frac{dF(\hat{\omega})}{d\hat{\omega}} = -4e^{-2\hat{\omega}}(e^{-2\hat{\omega}}+\sin{(2\hat{\omega})}-\cos{(2\hat{\omega})}).\label{eq:proof1}
\end{eqnarray}
$M(\hat{\omega}):=e^{-2\hat{\omega}}+\sin{(2\hat{\omega})}$ in Eq. (\ref{eq:proof1}) is minimized when $e^{-2\hat{\omega}}-\cos{(2\hat{\omega})}=0$ since the derivative of the function $M(\hat{\omega})$ is
\begin{eqnarray}
\frac{dM(\hat{\omega})}{d\hat{\omega}} = -2e^{-2\hat{\omega}}+2\cos{(2\hat{\omega})}.
\end{eqnarray}
Substituting $e^{-2\hat{\omega}}-\cos{(2\hat{\omega})}=0$ into Eq. (\ref{eq:proof1}), we have
\begin{eqnarray}
\frac{dF(\hat{\omega})}{d\hat{\omega}} \leq -4e^{-2\hat{\omega}}\sin{(2\hat{\omega})}<0
\end{eqnarray}
for $0<\hat{\omega}\leq\pi/4$. Since $F(0)=1$, Eq. (\ref{eq:state3}) is not violated. Thus, Eq. (\ref{eq:state2}) holds for all $\hat{\omega}\geq0$ implying that Eq. (\ref{eq:gaind}) is negative for $\hat{\omega}\geq0$. Therefore, the gain of the diffusion transfer function $|G(j\omega,L)|$ monotonically decreases for $\hat{\omega}\geq0$.

\smallskip
\par
The cut-off frequency, where the gain reaches -6 dB is uniquely determined since $|G(j\omega,L)|$ is a monotonically decreasing function. Calculating $20\log|G(j\omega,L)|=-6$ by binary search, we obtain $\hat{\omega}=1.44$ which leads to
\begin{eqnarray}
\omega_D = 4.14 \frac{\mu}{L^2}.\label{eq:cutoff}
\end{eqnarray}
Moreover, the condition of the communication distance $L$ for not further restricting the bandwidth of the signal output from the boundary layer by the diffusion layer, namely $\omega_D\geq\omega_B$, is determined by 
\begin{eqnarray}
L\leq2.03\sqrt{\frac{\mu}{\omega_B}}, \label{eq:relation}
\end{eqnarray}
where $\omega_B$ is the cut-off frequency of the boundary system. This can be an indicator to design the communication distance of MC channels if $\omega_B$ is known.

\smallskip
\par
It should be noted that the cut-off frequency $\omega_B$ is determined by the transmission system $G_1(s)$, the system $G_2(s)$, and the flux system $G_F(s,0)$. While the characteristics of the systems $G_1(s)$ and $G_2(s)$ are not specified, we know that $G_F(s,0)$ works like a differential element since $|G_F(s,0)|$ goes 0 when $s\rightarrow0$, and $|G_F(s,0)|$ goes $\infty$ when $s\rightarrow\infty$ in Eq. (\ref{eq:g2}). 

\smallskip
\par
Note that $G_B(s)$ can be decomposed into the transmission system $G_1(s)$ and the flux feedback system $G_{FF}(s)$, where
\begin{eqnarray}
G_{FF}(s) = \frac{1}{1-G_2(s)G_F(s,0)}.\label{eq:gff}
\end{eqnarray}
This decomposition is used in the next section to interpret $G_B(s)$ in more detail. Moreover, we define the cut-off frequency of $G_1(s)$ and $G_{FF}(s)$ as $\omega_1$ and $\omega_{FF}$, respectively.

\section{Numerical example}
\label{sec:numerical}

In this section, we consider an example of a specific MC channel with biological parameters and analyze its frequency response. We show using the example that the boundary system largely affects the communication bandwidth of the MC channel. In addition, we analyze the relation between the communication distance and the limitation of the bandwidth and verify Eq. (\ref{eq:relation}).

\smallskip
\par
Suppose the signal molecules are generated in a transmitter cell. In the boundary layer, the signal molecules are emitted from, or absorbed to the transmitter cell since they can go through the membrane freely with the rate constant of molecular transport $k$. Similarly, the signal molecules in the diffusion layer flows into/out the boundary layer. Thus, the dynamics in the boundary layer can be modeled by Eq. (\ref{eq:emission}), where the diffusion velocity $\hat{\mu}$ can be calculated by the Einstein–Smoluchowski relation \cite{Einstein1905}. Thus, the transfer functions $G_1(s)$ and $G_2(s)$ as
\begin{eqnarray}
    G_1(s) = \frac{k}{s+k}, \,\,\,\, G_2(s) = \frac{\hat{\mu}}{s+k}, \label{eq:tremission}
\end{eqnarray}
from which $G_B(s)$ is obtain as
\begin{eqnarray}
G_B(s) = \frac{k}{s+k-\hat{\mu}G_F(s)}.
\end{eqnarray}
In the diffusion layer, the signal molecules transmitted from the left boundary diffuse in a fluidic medium and are sensed by the receptors at the right boundary. Therefore, the dynamics in the diffusion layer can be modeled by the diffusion equation (\ref{eq:dif}) with the Dirichlet boundary condition $u(t,0)=v(t)$ at $r=0$ and the Neumann boundary condition Eq. (\ref{eq:neumann}) at $r=L$. Thus, the transfer function of the diffusion system $G_D(s,L)$ is obtained as Eq. (\ref{eq:g1}).

\smallskip
\par
We analyze the frequency response of the whole MC channel and the decomposed subsystems $G_1(s)$, $G_{FF}(s)$, and $G_D(s,L)$. We use the parameter sets shown in Table \ref{tab:param}. These parameter values are adopted from widely used parameters for numerical simulations of diffusion of molecules in synthetic biology.

\begin{table}
\begin{center}
\caption{Parameter sets for the analysis of the MC channel}
\label{tab:param}
\begin{tabular}{ccc}\hline\hline
Parameter&Value&Reference\\ \hline\hline
$k$&$5.0\times10^{-2}\,\si{s^{-1}}$&\cite{Pai2009}\\ 
$\mu$&$4.9\times10^{2}\,\si{\micro m^2\cdot  s^{-1}}$&\cite{Stewart2003}\\ 
$\hat{\mu}$&$9.9\,\si{\micro m\cdot  s^{-1}}$&\cite{Stewart2003,Einstein1905}\\ \hline\hline
\end{tabular}
\end{center}
\end{table}

\begin{figure}
    \centering
    \includegraphics[width=0.85\linewidth]{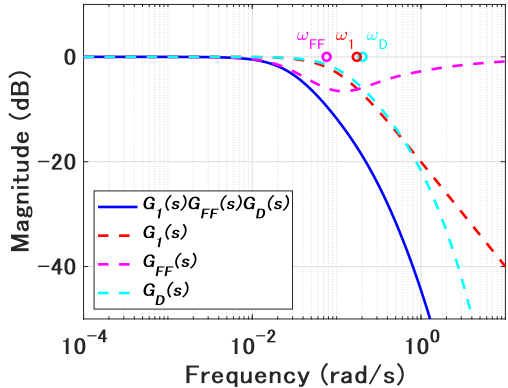}
    \caption{The gain diagram of the MC channel, the transmission system, the flux feedback system, and the diffusion system for $L=100\,\si{\micro m}$.}
    \label{fig:wgain}
\end{figure}

\smallskip
\par
Fig. \ref{fig:wgain} illustrates the gain of the MC channel, the transmission system $G_1(s)$, the flux feedback system $G_{FF}(s)$, and the diffusion system $G_D(s,L)$ when the communication distance is $L=100\,\si{\micro m}$. The gain of the MC channel (blue line) can be obtained by adding the gain of each subsystem together. Since the transmission system $G_1(s)$ is the first order lag system, the cut-off frequency of the transmission system is $\omega_1=\sqrt{3}k$. This implies that the bandwidth of the input signal to the membrane is initially restricted by the cut-off frequency $\omega_1$. Then, it passes through the membrane. The cut-off frequency of $G_D(s,L)$ defined in Section \ref{sec:limit} is $\omega_D=0.18$ rad/s. Since the cut-off frequency of the flux feedback system $\omega_{FF}(s)$ is the smallest cut-off frequencies in the subsystems, Fig. \ref{fig:wgain} shows that the dynamics in the boundary layer restricts the bandwidth the MC channel. This result implies that the dynamics in the boundary should be considered when one designs MC channels. 

\par
\smallskip
\noindent
{\bf Remark 1.} The gain of $G_{FF}(s)$ is basically 0 dB and have a dip around $\omega=0.06$ rad/s. The reason why the dip forms in the gain diagram of $G_{FF}(s)$ would be that $G_2(s)G_{F}(s)$ is the band-pass filter around $\omega=k$ because $G_2(s)$ is the low-pass filter and $G_{F}(s)$ works like a differential element. Since $G_{FF}(s)$ is the feedback system by the band-pass filter $G_2(s)G_{F}(s)$, the peak of the gain of $G_2(s)G_{F}(s)$ would causes the dip formation in the gain of $G_{FF}(s)$ around $\omega=k$. Therefore, it seems that the cut-off frequency $\omega_{FF}$ tends to be around $\omega=k$. 
 
\begin{figure}
    \centering
    \includegraphics[width=0.85\linewidth]{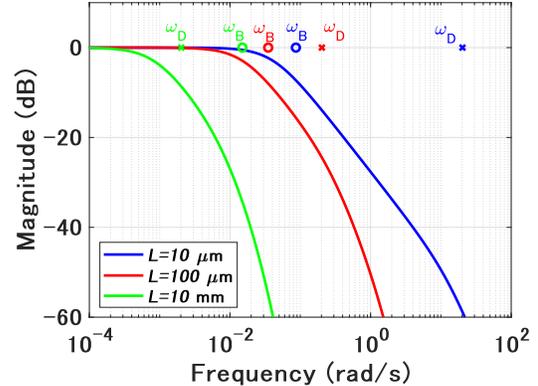}
    \caption{The gain diagram of the MC channel with the different communication distances.}
    \label{fig:Lcompare}
\end{figure}

\smallskip
\par
We have seen that the cut-off frequencies $\omega_1$ and $\omega_{FF}$ can restrict the bandwidth of the input signal to the boundary layer in Fig. \ref{fig:wgain}. It is necessary for the design to find the communication distance $L$ to avoid further bandwidth limitation by diffusion layer. Therefore, we analyze the frequency response of the MC channel for different communication distances $L=10\,\si{\micro m}$, $L=100\,\si{\micro m}$, and $L=1\,\si{mm}$ as shown in Fig. \ref{fig:Lcompare}. When $L=10\,\si{\micro m}$ and $L=100\,\si{\micro m}$, the cut-off frequency of the MC channel is determined by $\omega_B$ since the communication distance satisfies $L\leq2.03\sqrt{\mu/\omega_B}$. On the other hand, when $L=1\,\si{mm}$, $\omega_D$ is dominant and largely restricts the bandwidth comparing to $\omega_1$ and $\omega_{FF}$ since $L>2.03\sqrt{\mu/\omega_B}$. Thus, Fig. \ref{fig:Lcompare} shows that which subsystem determines the bandwidth of the MC channel, the diffusion system or the boundary system, depends on the distance, and Eq. (\ref{eq:relation}) can be an indicator of this relation.

\section{Conclusion}
\label{conc-sec}

In this paper, we have proposed a one-dimensional MC model with a boundary layer, and analyzed the frequency response of the proposed MC model based on the transfer functions. Moreover, we have shown the range of the communication distance $L$ that does not affect the bandwidth of the signal output from the boundary layer. Finally, by analyzing the frequency responses of a specific MC channel, we have shown that the boundary layer can restrict the bandwidth of the whole MC channel. 

\par
\smallskip
{\bf Acknowledgments: } The authors would like to thank Prof. Daisuke Tsubakino for helpful discussion for our theoretical development.
\appendices

\section{Derivation of the diffusion transfer function ($\ref{eq:rtrf}$) and the flux transfer function ($\ref{eq:rtrfder}$)}
\label{sec:dtf}

We derive the diffusion transfer function $G_D(s,r)$ and the flux transfer function $G_F(s,r)$ based on the calculation in \cite{Curtain2009}. By taking Laplace transform of Eq. (\ref{eq:dif}) for time $t$, we obtain
\begin{eqnarray}
su(s,r) = \mu \frac{\partial^2 u(s,r)}{\partial r^2}.
\label{eq:tlaplace}
\end{eqnarray}
Similarly, Lapace transform of Eq. (\ref{eq:tlaplace}) for the spatial variable $r$ leads to
\begin{eqnarray}
\frac{s}{\mu}u(s,p) = p^2 u(s,p)-pu(s,0) - \frac{\partial u(s,0)}{\partial r},
\label{eq:rlaplace}
\end{eqnarray}
where $p$ is the frequency variables for space. Eq. (\ref{eq:rlaplace}) can be decomposed into the partial fractions
\begin{eqnarray}
u(s,p) &=& \frac{1}{2}\left(\frac{1}{p+\sqrt{\frac{s}{\mu}}}+\frac{1}{p-\sqrt{\frac{s}{\mu}}}\right)u(s,0) \label{eq:pfd}\\
&&+ \frac{1}{2\sqrt{\frac{s}{\mu}}}\left(\frac{1}{p-\sqrt{\frac{s}{\mu}}}-\frac{1}{p+\sqrt{\frac{s}{\mu}}}\right)\frac{\partial u(s,0)}{\partial r}.\nonumber
\end{eqnarray}
Taking the inverse Laplace transformation for space, we have
\begin{eqnarray}
u(s,r) &=& \frac{1}{2}\left(e^{-\sqrt{\frac{s}{\mu}}r}+e^{\sqrt{\frac{s}{\mu}}r}\right)u(s,0) \label{eq:pfd2}\\
&&+ \frac{1}{2\sqrt{\frac{s}{\mu}}}\left(e^{\sqrt{\frac{s}{\mu}}r}-e^{-\sqrt{\frac{s}{\mu}}r}\right)\frac{\partial u(s,0)}{\partial r}.\nonumber
\end{eqnarray}
Since the derivative of $u(s,r)$ can be calculated as
\begin{eqnarray}
\frac{\partial u(s,r)}{\partial r} &=& \frac{\sqrt{\frac{s}{\mu}}}{2}\left(-e^{-\sqrt{\frac{s}{\mu}}r}+e^{\sqrt{\frac{s}{\mu}}r}\right)u(s,0) \label{eq:pfd3}\\
&&+ \frac{1}{2}\left(e^{\sqrt{\frac{s}{\mu}}r}+e^{-\sqrt{\frac{s}{\mu}}r}\right)\frac{\partial u(s,0)}{\partial r},\nonumber
\end{eqnarray}
by applying the Neumann boundary condition (\ref{eq:neumann}), we have 
\begin{eqnarray}
\frac{\partial u(s,0)}{\partial r} &=&\sqrt{\frac{s}{\mu}}\frac{e^{-\sqrt{\frac{s}{\mu}}L}-e^{\sqrt{\frac{s}{\mu}}L}}{e^{\sqrt{\frac{s}{\mu}}L}+e^{-\sqrt{\frac{s}{\mu}}L}}u(s,0). \label{eq:pfd4}
\end{eqnarray}
By substituting Eq. (\ref{eq:pfd4}) into Eq. (\ref{eq:pfd2}) and Eq. (\ref{eq:pfd3}), we obtain
\begin{eqnarray}
u(s,r) &=& G_D(s,r)u(s,0),\\
\frac{\partial u(s,r)}{\partial r} &=& G_F(s,r)u(s,0),
\label{eq:Ltrf}
\end{eqnarray}
where 
\begin{eqnarray}
G_D(s,r) &=& \frac{e^{-\frac{r}{\sqrt{\mu}}\sqrt{s}}+e^{\frac{r-2L}{\sqrt{\mu}}\sqrt{s}}}{1+e^{-2\frac{L}{\sqrt{\mu}}\sqrt{s}}},\\
G_F(s,r) &=& \sqrt{\frac{s}{\mu}}\frac{-e^{-\frac{r}{\sqrt{\mu}}\sqrt{s}}+e^{\frac{r-2L}{\sqrt{\mu}}\sqrt{s}}}{1+e^{-2\frac{L}{\sqrt{\mu}}\sqrt{s}}}.
\end{eqnarray}

\bibliography{IEEEabrv,bibdata}

\end{document}